\documentstyle[prl,aps,epsfig,tighten,twocolumn]{revtex}
\begin{document}
\draft
\title{Phase resolution limit in macroscopic interference between
	Bose-Einstein condensates}
\author{Sigmund Kohler and Fernando Sols}
\address{Departamento de F\'\i sica Te\'orica de la Materia Condensada 
	and Instituto ``Nicol\'as Cabrera'' \\
	Universidad Aut\'onoma de Madrid,
	E-28049 Madrid, Spain}
\date{November 5, 1999}
\maketitle
%
%-----------------------------------------------------------------------------
\begin{abstract}
We study the competition between phase definition and quantum phase
fluctuations in interference experiments between independently formed
Bose condensates.
While phase-sensitive detection of atoms makes the phase progressively
better defined, interactions tend to randomize it faster as
the uncertainty in the relative particle number grows.
A steady state is reached when the two effects cancel each other.
Then the phase resolution saturates to a value that grows with the
ratio between the interaction strength and the atom detection rate, and
the average phase and number begin to fluctuate classically.
We discuss how our study applies to both recently performed and
possible future experiments.

\end{abstract}

%-----------------------------------------------------------------------------

Recent macroscopic interference experiments \cite{andr97,ande98} have
revived interest on the question of whether two independent Bose
condensates can have a well-defined phase \cite{ande84,legg91}.
Guided by the experimental result
that a precise phase was indeed observed, the ensuing theoretical work
\cite{cast97,java96,cira96,nara96} has noted that, even if the initial relative
phase is random, as happens for pairs of condensates prepared in a Fock state,
the phase becomes progressively well-defined as atoms are emitted from the
compound system and recorded in a phase-sensitive detector. The resulting
view accords well with the standard quantum measurement picture according
to which the phase is created by the very act of measurement; if one tries
to measure the phase one indeed observes a definite phase.
However, these works have not discussed how the phase definition process
is affected by the mean-field
atomic interactions, which play a competing role by causing quantum
spreading of the phase.
In this article we study the role of interactions, and show that they
modify the above picture considerably. While the final conclusion
remains true that a well-defined phase is expected in a wide range of
experiments, we argue that this is possible only because the interactions
are sufficiently weak, and calculate explicitly how weak they must be in
a specific experimental setup. The central idea is that, due to interactions,
phase is not a constant of motion. An experiment will be effective in
measuring the phase only if the measurement is sufficiently intense
(in a sense that will be made precise) to overcome the phase internal dynamics
opposing definition.

We study the effect of interactions within the framework of the interference
experiment considered by Castin and Dalibard \cite{cast97}.
Two atom beams emitted from independent atom lasers \cite{bloc98} are made
to converge on an atom beam splitter \cite{adam94} and the emerging beams
are directed to two detectors. This experiment
has the attractive feature that, without being unrealistic, it lends
itself to a fundamental theoretical analysis, since it captures the essence
of phase measurement in its most simple form. With a good understanding of
the central physics, we will see that it is possible to draw some
semiquantitative conclusions on interference experiments, such
as that already performed at the MIT \cite{andr97}.

We consider two independent Bose condensates, each confined in a
harmonic trap. A Hamiltonian for the mean field description of the
interaction is easily derived from a Taylor expansion of the energy of
each condensate \cite{zapa98}.
We have $N_a$ ($N_b$) atoms in condensate $A$ ($B$), $N=N_a+N_b$ and
$n=N_a-N_b$. For $|n| \ll N$,
\begin{equation}
H = 2E(N/2) + E_{\rm c} n^2 /2
\label{hamiltonian}
\end{equation}
describes the coherent dynamics of the total system.
$E_{\rm c}=E''(N/2)$ determines to lowest order in $n$ the self-interaction
within each condensate, and can be derived from the ground state
properties of an interacting Bose gas \cite{dalf99}.
In a quantized version, the atom numbers $N_a$ and $N_b$ are given by the
operators $a^\dagger a$ and $b^\dagger b$.

Following Ref. \cite{cast97}, we consider an interference experiment in
which the condensate atoms are emitted from the traps and guided to a
50-50 beam splitter. The action of the two
detectors $D_{\pm}$ on the system is described by the operators
\begin{equation}
C_\pm = \sqrt{\gamma/2}\,(a\pm b),
\end{equation}
where $\gamma$ is the outcoupling rate per atom, and the expectation
values $I_\pm(t) = \langle\psi(t)|C_\pm^\dagger C_\pm|\psi(t)\rangle$
give the corresponding detection rates \cite{molm96}.

A convenient representation is provided by the phase states
\begin{equation}
\label{phasestates}
|\phi\rangle_N = {1 \over \sqrt{2^N N!}} \left(a^\dagger e^{i\phi}
+ b^\dagger e^{-i\phi}\right)^N |0,0\rangle .
\end{equation}
If the wave function of a state $|\psi\rangle$ in the
phase representation $\psi(\phi)= {}_N\langle\phi|\psi\rangle$ exhibits a sharp peak at an average value $\bar\phi$, it is easy to see that
the detection rates read $I_\pm=\gamma N [1\pm\cos(2\bar\phi)]/2$. This
property of full fringe visibility makes the phase states (\ref{phasestates})
ideal for interferometric purposes.
The number difference operator $n=a^\dagger a-b^\dagger b$ acts on 
the phase states as
$n|\phi\rangle_N = -i(\partial/\partial\phi)|\phi\rangle_N$.
Thus $n$ and $\phi$ are conjugate quantities, and, for a sharp phase
distribution, the inequality $\Delta n\,\Delta\phi \geq 1/2$ applies.
The Heisenberg equations of motion
$\partial\phi/\partial t=-E_{\rm c}n/\hbar$, $\partial n/\partial t=0$,
lead to a time-dependent interference pattern
$I_\pm(t) \propto 1\pm\cos(2E_{\rm c} \bar{n} t/\hbar)$.

A necessary feature of phase-sensitive detection is that it does not give
information on which condensate the recorded atom comes from. Each detected
atom may come with equal probability from either condensate $A$ or $B$.
As a result, the uncertainty in the relative atom number after $k$ detections grows like $\Delta n = \sqrt{k}$. In the absence of interactions \cite{cast97}, 
a minimal wave packet is formed, so that $\Delta\phi = 1/2\sqrt{k}$. Let us
assume that the same result holds true in the presence of interactions. Then we can conclude that the increase $\Delta n^2\to\Delta n^2+1$ in the number uncertainty occurring in each additional detection, is accompanied by a corresponding decrease in the phase uncertainty,
\begin{equation}
\label{decrease}
\Delta\phi \to {1 \over 2\sqrt{\Delta n^2 + 1}}
\approx \Delta\phi - 2\Delta\phi^3 .
\end{equation}
On the other hand, interactions generate phase dynamics.
The mean value of the phase shows a drift,
$\bar{\phi}(t)=\bar{\phi}(0)-E_{\rm c} \bar n t/\hbar$, while
its variance undergoes ballistic spreading,
$\Delta\phi(t)=\Delta\phi(0) + E_{\rm c}\Delta n \,t/\hbar$. Thus between
detections the phase uncertainty grows at a rate proportional to $\Delta n$,
which in turn grows with the number of detections. Since $\gamma N$ is the
total detection rate, the mean increase of phase variance between detections is 
\begin{equation}
\label{increase}
\Delta\phi \to \Delta\phi + {E_{\rm c} \over 2\hbar\gamma N\Delta\phi } .
\end{equation}
In the early stages of the measurement process $\Delta n$ is small and the
effect of interactions is negligible; the variation in (\ref{increase}) is
much smaller than that in (\ref{decrease}). However, as $k$ grows, the two
effects become comparable and cancel each other.
Then a steady state is formed in which {\it the phase becomes defined by the
detections at the same rate at which it is randomized by the interactions\/}.
The stationary values of the variances are
\begin{equation} \label{eq:steadystate}
\Delta\phi_s = \left({\kappa\over 4}\right)^{1/4}, \quad
\Delta n_s = \left({1\over 4\kappa}\right)^{1/4}, \quad
\kappa \equiv\frac{E_{\rm c}}{\hbar\gamma N}.
\end{equation}
Saturation occurs after $k^*=\Delta n_s^2$ atoms have been detected. 
In the absence of interactions,
$\Delta\phi_s = 0$, and the phase can become arbitrarily well-defined.
In contrast to that, Eq.~(\ref{eq:steadystate}) indicates that {\it interactions cause an intrinsic limitation in the achievable phase resolution\/}.
A steady state with a well-defined phase $\Delta\phi\ll 2\pi$
can be established only if interactions are sufficiently weak ($\kappa\ll 1$).
For a 3d harmonic trap in the Thomas-Fermi limit, $E_{\rm c}\sim N^{-3/5}$
\cite{zapa98,dalf99,baym96} and thus the effective interaction parameter,
$\kappa \sim N^{-8/5}$, increases during the detection process. However,
we shall focus on time scales where $N$ is practically constant ($k\ll N$).

For a more quantitative analysis we have performed a numerical simulation of 
the time evolution in a single experiment based on the quantum jump method \cite{molm96}.
Starting from an inital Fock state $|\psi(0)\rangle=|N_a,N_b\rangle$, the
time evolution over a sufficiently small time interval $\Delta t$ can be
simulated as follows.
If an atom is detected at $D_\pm$, the wave function changes as
\begin{equation} 
\label{eq:detection}
|\psi(t)\rangle \to
C_\pm|\psi(t)\rangle.
\end{equation}
These detections occur with probabilities
\begin{equation}
p_\pm(t,\Delta t)
= \langle\psi(t)|C_\pm^\dagger C_\pm|\psi(t)\rangle \Delta t \ll 1.
\end{equation}
If no detection takes place, the system propagates from $t$ to $t+\Delta t$
with the effective Hamiltonian
\begin{displaymath}
H_{\rm eff} 
= H -i (C_+^\dagger C_+ + C_-^\dagger C_-)/2
= H -i\gamma (a^\dagger a + b^\dagger b)/2 .
\end{displaymath}
Since the propagation with $H_{\rm eff}$ as well as the detection
(\ref{eq:detection}) is non-unitary, the wave function has to be normalized
after each time step.
In the present case, an important simplification is brought about by the fact that the
time evolution with $H_{\rm eff}$ leaves the detection probabilities unchanged.
This means that the detection times are uncorrelated, and the time $t$
between consecutive detections is exponentially distributed,
$w(t)=\gamma N\exp(-\gamma N t)$ being its probability distribution.
Therefore the stochastic time evolution described above is equivalent
to one which can be implemented more efficiently, namely that
the system propagates with the Hamiltonian (\ref{hamiltonian}) over an
exponentially distributed random time $t$ and after each propagation a
detection (\ref{eq:detection}) is performed at either $D_+$ or $D_-$ with
probability $P_\pm(t) = I_\pm/(I_+ + I_-)$.

\begin{figure}
\centerline{ \psfig{width=8.0cm,figure=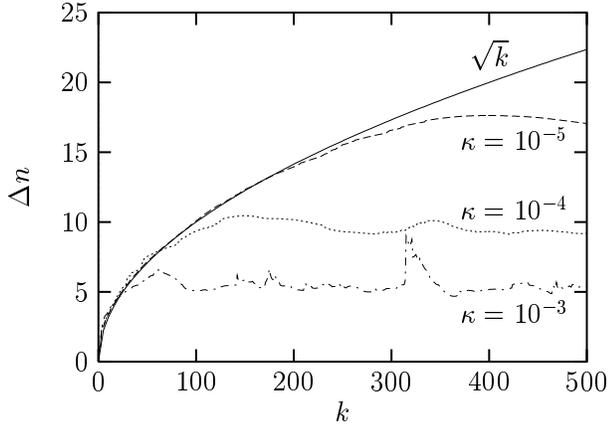} }
\vspace*{3ex}
\caption{
Time evolution of the number variance $\Delta n$ in a typical quantum
trajectory for different interaction strengths
$\kappa\equiv E_{\rm c}/\hbar\gamma N$.
The full line depicts the behavior in absence of interaction.
Initially, the total atom number is $N=10^6$ and the number
difference is $n_0=200$.
\label{fig:var_n}
}
\end{figure}

\begin{figure}
\centerline{ \psfig{width=7.0cm,figure=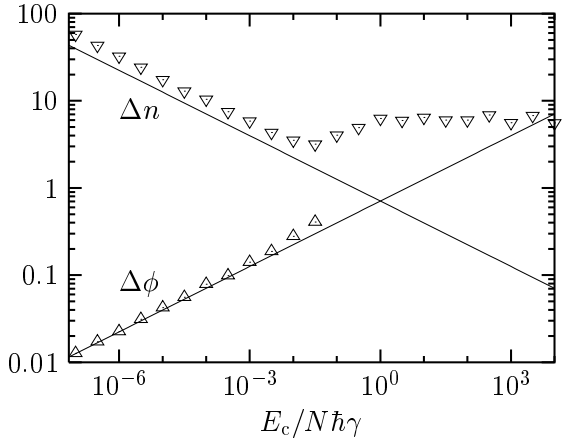} }
\vspace*{3ex}
\caption{
Stationary variances $\Delta n_s$ and $\Delta\phi_s$ for different
interaction strength.
The average values in the steady state (dots) have been calculated from 20
simulations with initially $N=10^6$ atoms. They are in good agreement with
the analytical estimates in Eq.~(\ref{eq:steadystate}) (full lines).
We do not plot $\Delta\phi_s$ for large values of $\kappa$ because it is not
well defined once it becomes of order unity.
\label{fig:steadyvar}
}
\end{figure}

\begin{figure}
\centerline{ \psfig{width=7.8cm,figure=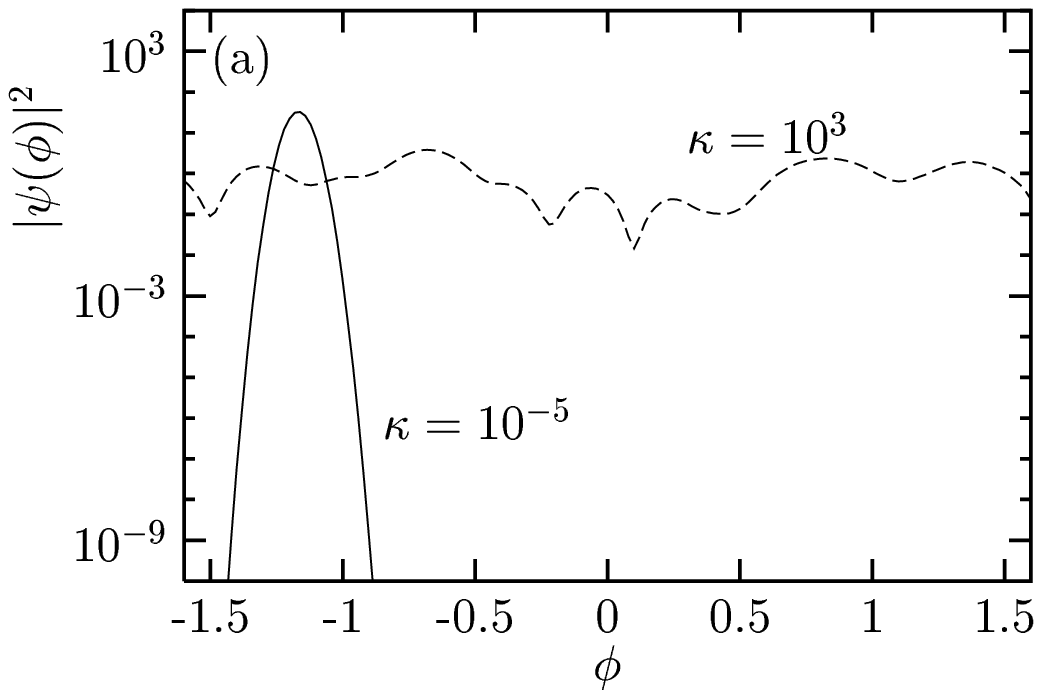} }
\vspace*{3ex}
\centerline{ \psfig{width=7.5cm,figure=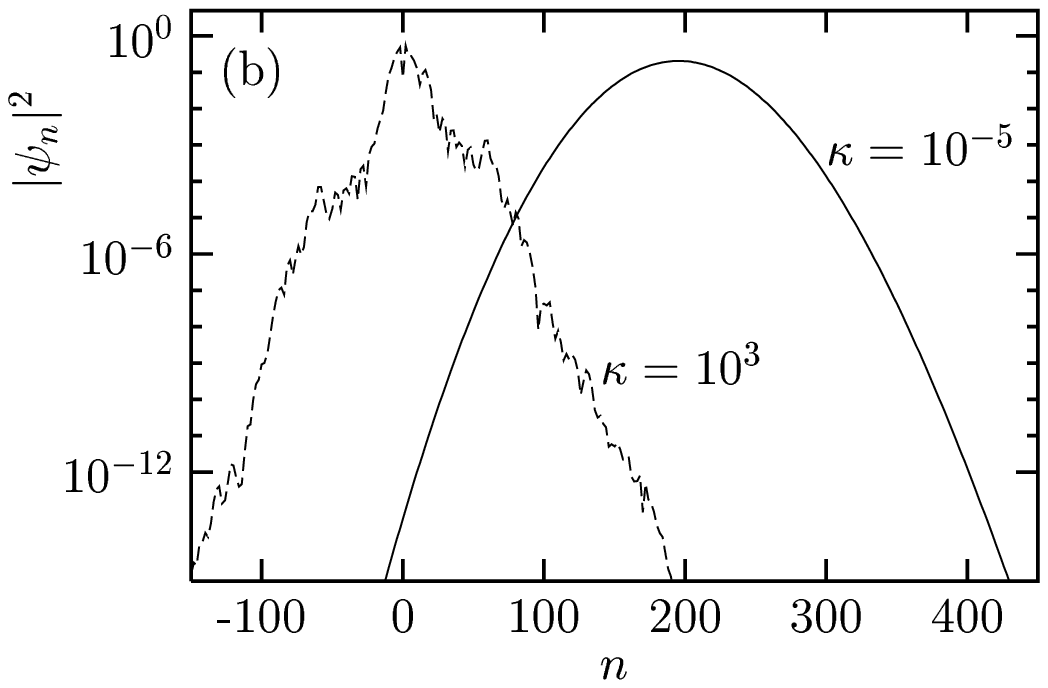} }
\vspace*{3ex}
\caption{Phase (a) and number (b) distribution after 500 detections for
weak (full lines, $n_0=200$) and strong (dashed, $n_0=0$) interaction
for $N=10^6$.
\label{fig:wavefunction}
}
\end{figure}

Figure~\ref{fig:var_n} shows how the number variance evolves in a typical experimental run for different interaction strengths.
Starting the stochastic time evolution with the initial state
$|N_a,N_b\rangle$, the variance $\Delta n$ grows with the
square root of the number of detected atoms as in the interaction free case
\cite{cast97}.
After $k^*$ atoms have been detected, {\it i.e.} after a time
$t^*=(4N\gamma E_{\rm c}/\hbar)^{-1/2}$, the interaction becomes effective, and
the variances saturate at the values (\ref{eq:steadystate}).
Figure~\ref{fig:steadyvar} shows the obtained $\Delta n_s$ and $\Delta\phi_s$
as a function of the interaction strength.
For weak interactions, our numerical studies confirm to a good approximation
the estimates (\ref{eq:steadystate}).
We also find good agreement with our asumption of a gaussian shape for
the wavefunction at large $k$ (see Fig.~\ref{fig:wavefunction}).

Figure \ref{fig:steadyvar} also reveals that for strong interactions
($\kappa \gg 1$) $\Delta n_s$ saturates as a function of $\kappa$.
In this case, we expect the phase to become
completely random. This corresponds to the limit in which interactions
dominate over phase-sensitive detection.
Figure \ref{fig:wavefunction}a shows, indeed, that in this case, instead of
a gaussian phase distribution we find asymptotically a completely undefined
phase. On the other hand, the
random variation of $n$ by $\pm 1$ makes each detection act like a random kick,
in analogy to the kicked rotor problem \cite{casa79} with $n$ playing the
role of the angular momentum.
This dynamical system is known to have a regime where the angle distribution
spreads over the circle at long times, while the angular momentum
distribution becomes exponentially localized within a length that
depends only on the kick strength, and not on the parameters of the
free propagation \cite{casa79}.
Thus we expect to find wave functions of the form
$\psi_n \sim\exp(-|n-\bar n|/4\Delta n_s)$, where $\Delta n_s$ is independent
of $\kappa$. This is confirmed in Figs.~\ref{fig:steadyvar}
and \ref{fig:wavefunction}b.

In the non-interacting case \cite{cast97} each detection amounts to a
partial measurement of the phase, the measurement becoming more precise as
more atoms are recorded. However, what is actually measured is $\cos(2\phi)$
through the observed intensities $I_\pm$. Thus the whole measurement process
determines the phase only up to a sign ($-\pi/2\leq\phi<\pi/2$), and the
condensate is in a
superposition of two minimal wave packets with peaks at $\pm\bar\phi$.
Within each peak $\Delta n=\sqrt{k}$ and $\Delta\phi=(2\Delta n)^{-1}$.

\begin{figure}
\centerline{ \psfig{width=7.5cm,figure=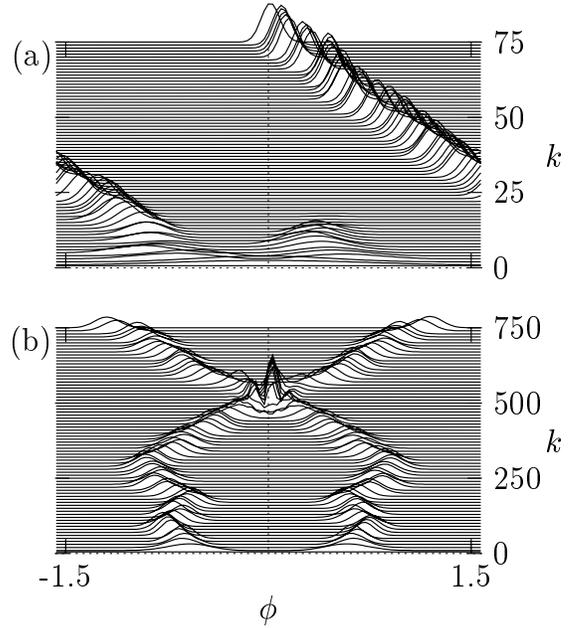} }
\vspace*{3ex}
\caption{
Time evolution of the phase distribution $|\psi(\phi)|^2$ for $\kappa=2\times
10^{-4}$ and $N=10^5$. The initial number differences are $n_0=200$ (a) and
$n_0=1$ (b).
For graphical reasons, we plot in panel (b) only after every tenth detection.
\label{fig:phase}
}
\end{figure}

To study the effect of interactions on the phase definition process, let us 
assume first that the initial number difference is
$n_0\sim\sqrt{N}\gg\Delta n_s$.
Figure \ref{fig:phase}a shows that, in contrast to the interaction free
case, after an initial transient the system evolves into a single
wave packet with an average phase $\bar\phi=E_{\rm c}\bar n t/\hbar$.
The reason is the following. Since $\bar n\approx n_0$, the sign of $\bar n$
is known. On the other hand, the detection record yields both, $I_\pm(t)\sim
\cos(2\bar\phi)$ and $\dot I_\pm(t)\sim \bar n\sin(2\bar\phi)$.
This makes the measurement of $\phi$ unambigous \cite{comm2,zure93}.

When $n_0 \alt \Delta n_s$, the sign of $\bar n$ cannot be measured with enough
precision and, as in the non-interacting case, $\bar\phi$ is determined
from the measurement only up to a sign. Accordingly, the phase distribution
evolves into two wave packets. The sign of $n$ is correlated with ${\phi}$
in such a way that the sign of $n\sin(2\phi)$ is the same for both components.
This can be seen in Fig.~\ref{fig:phase}b, where the two wave packets, each
with an opposite mean phase value, drift with opposite velocity.

Inspection of Fig.~\ref{fig:phase}a seems to suggest that $\bar{n}$ stays
steady once $\Delta \phi$ and $\Delta n$ reach saturation. However, a
closer look reveals that this is not the case.
We find that $\bar{n}$ remains almost constant during the initial interval
$k<k^*$ (not shown). Once $\Delta n$ saturates to $\Delta n_s$ for $k>k^*$,
$\bar{n}$ begins to fluctuate classically with the spread over many runs growing
like $\sqrt{k}$. This implies that the total number uncertainty obtained by
averaging the quantum averages $\bar{n}$ over many runs grows steadily like
$\sqrt{k}$. This should be expected, since the result of such a combined
average must be identical to the full quantum average of two condensates
emitting at random.

To understand the MIT experiment \cite{andr97}, we may identify the emitted
atoms with those in the region where the two wave packets overlap.
Then we may replace the outcoupling rate per atom $\gamma$ by the rate
of overlap increase, which is of order $10\,{\rm s}^{-1}$. Since
$E_{\rm c}/\hbar\approx 0.01\,{\rm s}^{-1}$  for $N\approx 5\times10^6$ sodium
atoms, this results in a phase resolution $\Delta\phi_s \approx 3\times
10^{-3}$, which is consistent with the observation of a clear interference
pattern \cite{andr97}.

In summary, we have studied the influence of self-interaction on the phase
definition in a macroscopic interference experiment between separate Bose
condensates. Our main finding is that the ratio between interaction strength
and detection rate limits the phase resolution. This arises when the
phase randomizes due to interactions at the same rate at which it becomes
defined by the successive partial measurements performed at each atom
detection. We have derived an analytical expression for the phase resolution
which has been confirmed by a Monte Carlo simulation.
Once the interference pattern reaches its intrinsic resolution limit,
the average phase and number fluctuate classically.
The studied model allows us to understand in simple terms why, despite
interactions, good phase resolution can be obtained in macroscopic
interference between Bose condensates.
It will be most desirable to perform experiments that explore the
fading of the interference pattern with increasing interaction
strength.

We wish to thank C.~Bruder and W.~Ketterle for valuable discussions.
This work has been supported by the EU TMR Programme under Contract No.\
FMRX-CT96-0042 and by the Direcci\'on General de
Investigaci\'on Cient\'{\i}fica y T\'ecnica under Grant No.\ PB96-0080-C02.

%\end{document}
%
%
%-----------------------------------------------------------------------------
%%%
%%%  the figures
%%%
%-----------
%-----------
%-----------
%-----------


\begin{thebibliography}{19}
\bibitem{comm2}
An interesting corollary is that a set of sequential partial measurements
of the phase $\phi$ yields a partial measurement of its time derivative
$\dot{\phi}$, and hence of its canonical conjugate $n$.
The compromise is the formation of a generalized coherent state in phase
and number. This conclusion is consistent with the ubiquity in nature of 
coherent states behaving like classical entities.
We note that these coherent states do not require the presence of a 
harmonic confining potential as assumed in Ref.~\cite{zure93}.
\end{thebibliography}
\end{document}